\def\fleck{P1}
\def\kadek{P2}
\def\aimee{P3}
\def\beck{P4}
\def\sr{P5}
\def\ap{P6}
\documentclass{vgtc}                          % final (conference style)
\ifpdf%                                % if we use pdflatex
  \pdfoutput=1\relax                   % create PDFs from pdfLaTeX
  \usepackage{graphicx}                % allow us to embed graphics files
  \DeclareGraphicsExtensions{.pdf,.png,.jpg,.jpeg} % for pdflatex we expect .pdf, .png, or .jpg files
\else%                                 % else we use pure latex
  \usepackage{graphicx}                % allow us to embed graphics files
  \DeclareGraphicsExtensions{.eps}     % for pure latex we expect eps files
\fi%

\graphicspath{{figures/}{pictures/}{images/}{./}} % where to search for the images

\usepackage{microtype}                 % use micro-typography (slightly more compact, better to read)
\PassOptionsToPackage{warn}{textcomp}  % to address font issues with \textrightarrow
\usepackage{textcomp}                  % use better special symbols
\usepackage{mathptmx}                  % use matching math font
\usepackage{times}                     % we use Times as the main font
\usepackage{cite}                      % needed to automatically sort the references
\usepackage{tabu}                      % only used for the table example
\usepackage{booktabs}                  % only used for the table example
\onlineid{0}

\vgtccategory{Research}

\vgtcinsertpkg

\title{Designing Situated Dashboards: Challenges and Opportunities}

\author{Anika Sayara\thanks{e-mail: sayanika@cs.ubc.ca}\\ %
        \scriptsize University of British Columbia %
\and Benjamin Lee\thanks{e-mail: Benjamin.Lee@visus.uni-stuttgart.de}\\ %
     \scriptsize University of Stuttgart %
\and Carlos Quijano-Chavez\thanks{e-mail: quijancr@visus.uni-stuttgart.de}\\ %
     \scriptsize University of Stuttgart %
\and Michael Sedlmair\thanks{e-mail: Michael.Sedlmair@visus.uni-stuttgart.de}\\ %
     \scriptsize University of Stuttgart} %

\teaser{
  \centering
  \makebox[\textwidth][c]{\includegraphics[width=1.125\linewidth]{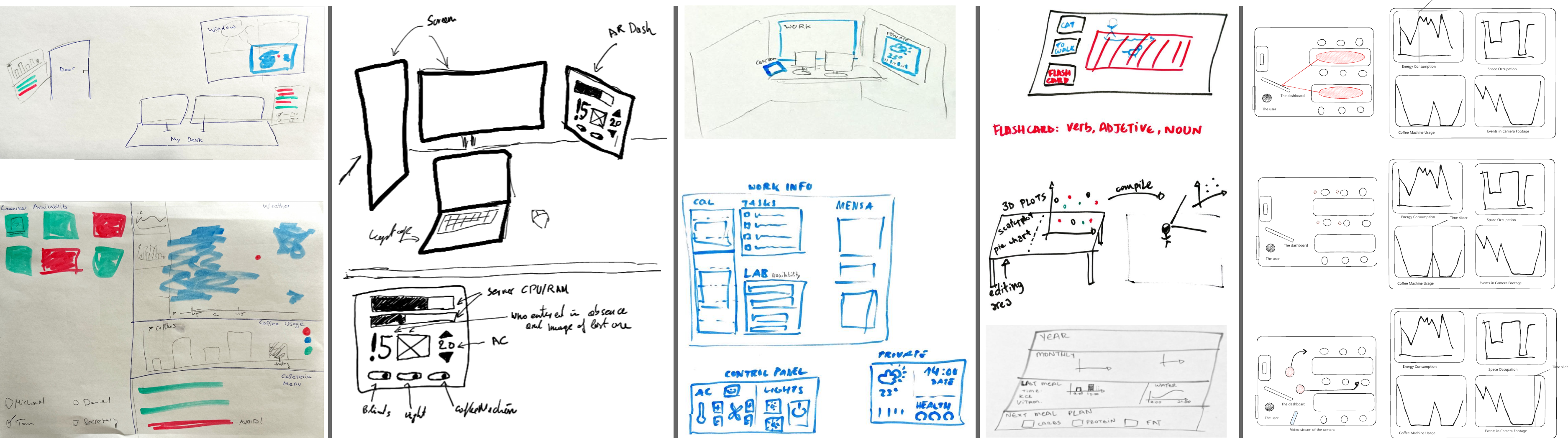}}
  \vspace{-6mm}
  \caption{
  Situated dashboard sketching scenarios of participants. From left to right: \beck{} sketched multiple panels scattered around the office close to relevant subjects (top) and a single panel (bottom); \fleck{} used situated dashboards to control physical referents: air-conditioner, blinds, etc.; \sr{} grouped categories of information in panels (bottom) and placed them around the room (top); \aimee{} sketched three tabletop-situated dashboards; \ap{} used situated dashboard to playback time series data as situated visualizations: energy consumption as colored highlights (top), coffee consumption as number of cups (middle), and playback of camera footage via holograms (bottom). \kadek{} decided not to sketch.
  }
  \label{fig:teaser}
}

\abstract{
Situated Visualization is an emerging field that unites several areas - visualization, augmented reality, human-computer interaction, and internet-of-things, to support human data activities within the ubiquitous world. Likewise, dashboards are broadly used to simplify complex data through multiple views. However, dashboards are only adapted for desktop settings, and requires visual strategies to support situatedness. We propose the concept of AR-based situated dashboards and present design considerations and challenges developed over interviews with experts. These challenges aim to propose directions and opportunities for facilitating the effective designing and authoring of situated dashboards.
} % end of abstract

\CCScatlist{
  Situated Dashboard, Dashboard Design, Situated Analytics, Data Visualization, Augmented Reality
}

\begin{document}


%\firstsection{Introduction}

\maketitle
\section{Introduction}
Data is ubiquitous in the physical world around us. A person may desire to understand more about some passive referent, or to keep informed of the state of some actively updating referent \cite{fleckRagRugToolkitSituated2022}. With augmented reality (AR) head-mounted displays (HMD), it is possible to decrease the level of spatial indirection between the referent and its data, such that it is displayed close to or even embedded on top of each other \cite{willettEmbeddedDataRepresentations2017}. Thus, many works on situated visualization have sought to minimize this indirection, whether it be to overlay AR visualizations directly on top of grocery store products \cite{elsayedSituatedAnalytics2015}, display information about a building next to it \cite{reitmayrCollaborativeAugmentedReality2004}, or show temporal data next to temperature sensors in a building \cite{fleckRagRugToolkitSituated2022}.

In the physical world, we are bound by physical constraints. In particular, the design of situated visualization is influenced by its navigational requirements  \cite{leeDesignPatternsSituated2023}. For example, if the physical referents are spread across a large area, the use of embedded visualizations may be problematic due to the physical and mental effort required to locate and navigate to the referents. Thus, using a \textit{many-to-one} view may help consolidate such spatially distributed information into a singular visual representation \cite{leeDesignPatternsSituated2023}.

In traditional desktop computing, visualization dashboards are vital in their ability to also consolidate large amounts of disparate information into a format that provides an overview of the data. Dashboards are particularly useful for hiding the complexities of the logical world from end-users, making data easily accessible without the end-user needing to know where and how the data comes from.

Therefore, we propose the concept of AR-based \textit{situated dashboards}. While viewing the data in its exact physical context can be useful, situated dashboards may accommodate situations where needing to be in said physical context is too cumbersome or impractical. For example, an AR-situated dashboard might provide a factory manager with the status of all operations on the floor at any moment's notice. The application could then help the manager navigate to a problematic areas of the factory (e.g., using situated AR navigational instructions \cite{reitmayrCollaborativeAugmentedReality2004,mulloniHandheldAugmentedReality2011}). The dashboard may then transition into an embedded view \cite{willettEmbeddedDataRepresentations2017} for the manager to engage in problem-solving within the physical context.

While traditional visualization dashboards are commonplace, and are technically already in use in many situated contexts, there has been little to no exploration on the use of situated dashboards in AR, which is now arguably the de-facto standard for situated visualization \cite{bressaWhatSituationSituated2022,shinRealitySituationSurvey2023}. The possibilities of situated dashboards are vast, and there is no clear definition or approach for how they can be designed, created, or even evaluated. In this position paper, we establish a preliminary understanding of situated dashboards through a set of six interviews with researchers in both situated visualization and AR. Our interviews focus on understanding experts' perception about situated dashboards, design considerations, and potential challenges of designing and authoring situated dashboards.

\section{Related Work}

%Many authoring AR studies have been proposed in the HCI community focusing on simulating large-scale subjects (e.g. \cite{CSpace2020, constructingBuildingMixedReality2023}), context-aware experiences (e.g. ), portable interfaces, and proxemic-gestural interactions.
%Our scope in this work is mainly to explore prototyping dashboard design, situated visualization, and authoring immersive visualization tools.
Since our study is focused on assessing dashboard design for situated visualization, we surveyed the literature on: 
(1)  dashboard design considerations and (2) situated visualizations. 
%and (3) authoring tools using augmented reality in situated scenarios.

\subsection{Dashboard Design}
\label{sec:dashboardDesign}
Dashboards are broadly used in business intelligence to support users in analyzing complex data sets through multiple views~\cite{multipleViewsBaldonado2000} and the coordination between them~\cite{scherr2008multiple}.  Dashboard design guidelines emerged to advise visual perception, information load, and interactions (e.g.~\cite{few2006information,few2007dashboard,rasmussen2009business,kitchin2015knowing,BUGWANDEEN2019,QualDash2020,AnsweingChallengesEmergencyResponses2022}). Popular visualization tools like Tableau and PowerBI contain huge galleries of templates in order to generate dashboards. However, such systems are challenging to use for non-experts. On the other hand, researchers make efforts to provide authoring and visualization recommendations tools (e.g.~\cite{GuidedMultiView2021,multivision2022,medley2023}). 
Recently Bach et al.~\cite{Bach2023DashboardDesignPatterns} surveyed dashboard designs and detailed 48 design patterns. They mapped solutions: \textit{data abstraction, screenspace organizing, grouping of elements, relations encoding and the interaction or personalization} in the dashboard design process. Despite those studies being focused on conventional displays, we considered those solutions to prepare the questions.

\subsection{Situated Visualization}

Willett et al.~\cite{willettEmbeddedDataRepresentations2017} defined \textit{situated visualizations} as a situated data representation in a relevant location where the representations are connected to physical referents. When referents are not accessible, referents can be represented using scaled 3D models (\textit{proxies})~\cite{proxSituated2023}. 
%The research community considers the importance of \textit{situatedness} and \textit{visualization} in its studies and has a common interest in bringing closer visualizations into people's everyday environments~\cite{bressaWhatSituationSituated2022}. 
Bressa et al.~\cite{bressaWhatSituationSituated2022} surveyed studies and proposed perspectives to categorize the concept of situatedness: (1) \textit{space} puts emphasis on the spatial organization and relationship between the physical environment and visualizations; (2) \textit{time} focuses on the distance in time between the gathered data and its presentation; (3) \textit{place} considers the meaningful location where users act; (4) \textit{activity} refers to the human activities that designers need to consider with visualizations being appropriated to contexts; and (5) \textit{community} emphasizes in the audience, i.e., designers and developers.
%who are designers of visualizations and look for sharing and supporting issues. 
Each perspective opens challenges and motives our intention to design dashboards. 

Recently, \textit{active proxy dashboard} was proposed to analyze abstract visualizations from \textit{proxies} through tangible interactions~\cite{ActiveProxy2023}. The main idea is to build binding events between proxies and data representations, allowing analysts to interact directly with proxies and visualizations that are displayed on conventional screens. Although the advantages of analyzing inaccessible referents and using powerful known displays, limitations about \textit{place} and \textit{activity} perspectives emerged. The context-dependent from human activities relies on context recreation difficulties. We believe that authoring tools will be closer to creating dashboards context-independent.

Furthermore, multiple studies seek to standardize properties and to establish guidelines that mitigate the challenges of multiple situated views. Batch et al.~\cite{ViewManagament2023} evaluated different ways of view management and identified properties to consider in future implementations. 
%studied the use of a shadowbox, world-in-miniature, and guided-tour to evaluate different ways of view management. In addition, their study identified properties to consider in future implementations.
More formally, Lee et al.~\cite{leeDesignPatternsSituated2023} identified patterns, dimensions, and guidelines on how to investigate situated visualization.

\section{Interview Methodology}%Study Design
A main contribution of our work is the results from semi-structured expert interviews of six AR and/or visualization researchers. The interviews aim to characterize challenges and opportunities for situated dashboard design. A focus group approach was not considered due to timeline constraints.

The participants were recruited through convenience sampling, and had varying levels of expertise in AR, data visualization, and situated visualization. Four of the six participants have published at least one paper on situated visualization/analytics.
Three participants were interviewed in person, and the other three were interviewed remotely. The session started with the participants describing their perception of situated dashboards. They were then tasked with ideating an AR HMD based situated dashboard for their typical workday at the office. At this time, the in-person participants were provided with pen and paper and the remote participants with Excalidraw board. While most participants used these, one remote participant chose not to sketch during the session but later emailed us a sketch. Another remote participant only described their ideas verbally. Throughout this process, participants were encouraged to articulate their thoughts regarding various aspects of their designs, including features, context, interactions, user experience, and potential implementation challenges. During this time the participants were also asked to reflect on their past experiences and talk about workflows for implementing situated visualizations. 
Following a reflexive thematic analysis method \cite{braun2019reflecting}, we collected, transcribed, and analyzed the interview data.

\section{Design Considerations of Situated Dashboards}
Our participants' perceptions of situated dashboards and their design considerations were fragmented, with diverse and sometimes conflicting views. Motivated by solutions proposed in the literature (Section~\ref{sec:dashboardDesign}), we discuss five main considerations:
(1) What content do situated dashboards display;
(2) What do they look like;
(3) Where are they situated;
(4) What interactions do they facilitate; and
(5) How can they be customized?

% Participants’ perception about situated dashboards appear to be fragmented, with diverse and sometimes conflicting views about - (1) What should be displayed? \textit{(content)}, (2) What should it look like? \textit{(visual features)}, 3) Where should it be situated? \textit{(situatedness)}, 4) What sort of interaction should it have \textit{(interaction)}, and 5) What sort of customization should it support to facilitate personalization? \textit{(customization)}

% In this section, we discuss these aspects of situated dashboards as design considerations.
%In this section, we show the findings found as design considerations.
\subsection{Content of Situated Dashboards}
When asked what they thought ``dashboards'' meant, most participants associated the term with data visualization. According to \beck{}, \fleck{}, \sr{}, and \kadek{}, a dashboard contains information from multiple sources that serves as an overview or summary. \aimee{} identified this overview as the differentiating factor between situated visualization and a situated dashboard.
However, they also defined a dashboard not by its content but by its ability to control something. Thus, a dashboard is \textit{``a surface where we will be able to control something.''} By this definition, seeing the data alone is not sufficient to be considered a dashboard.

For ``situated dashboards'', \ap{} preferred a narrower definition. In their words: \textit{``a dashboard needs to have more than one visualization about the same group of elements of physical objects. So when I hear a situated dashboard, it would mean that there are at least two visualizations about a specific object or a specific part of the physical workspace.''}
\sr{} provided some considerations about the selection of content for a situated dashboard. According to them, the information shown on a situated dashboard depends on: (1) where the dashboard is placed in the environment; (2) who the user is; and (3) what information is important to the user in a target situation. Another important consideration, as pointed out by \sr{}, is the privacy of the information presented on the dashboard, particularly if the situated dashboard is placed in a shared context.

\subsection{Appearance of Situated Dashboards}
When asked to design a situated dashboard,  all participants laid out their visualizations on a rectangular 2D space (Figure~\ref{fig:teaser}). \fleck{}, \aimee{}, and \ap{} used a single panel to display all information. \sr{} instead used multiple panels, with each containing a group of relevant information. These panels were then scattered near their referents. \beck{} designed for both scenarios. When using a single panel for displaying multiple data sources, \beck{} designed their dashboard in such a way that the layout of the dashboard changes dynamically to focus on the information that is most relevant to the location or task of the user.

\ap{} went even further by creating associations between individual visualizations on their dashboard and physical referents through visual links (red lines on \ap{}'s sketch in Figure~\ref{fig:teaser}). Their intention with this was to \textit{``to see the data streaming from the physical referent to the dashboard to the visualization itself.''}

%\begin{figure}[t]
%\centering
%\includegraphics[width=\linewidth]{figures/APdrawing.pdf}
%\vspace{-9mm}
%\caption{Sketch of situated dashboards by \ap: (a) using the slider on the energy consumption visualization of the situated dashboard to move in time and see the level of energy consumption through colors of the situated visualization [in the picture represented in red]. The red line links the visualization of the dashboard to situated visualization. (b) moving slider to see coffee consumption represented as number of cups (brown circles) next to the person (c) use dashboard to slide through camera footage and watch it play out as situated visualization and also in the video screen of the camera. 
%}
%\label{fig:Prouzeau1}
%\end{figure}

\subsection{Situatedness of the Dashboard}
According to \ap{}, \textit{``you don't situate the dashboard, you situate the information from the dashboard.''} All participants agreed that a situated dashboard should be in a specific context where it provides relevant and actionable information to users. As an example, \ap{} suggested that situated dashboards could be placed in the environment in order to playback time series data as situated visualizations.

\kadek{} pointed out that the contents of a dashboard could be situated either near a physical referent, or near a tangible or virtual proxy of said referent \cite{proxSituated2023}. They gave the example of a factory manager having a situated dashboard which was positioned near a tangible or virtual model of factory machinery in their office.
\aimee{} cautioned against the embedding of dashboards directly onto referents, stating \textit{``I would have a hard time situating a dashboard into a very specific object, because if a dashboard is a lot of info why would I want to stick that to one object, unless that object was related to all the information.''} As a possible workaround, \beck{} suggested that the dashboard's layout could dynamically change according to the user's context in order to minimize clutter. For example, during lunchtime, food-related data would mainly be shown on the dashboard with everything else being minimized. During work hours however, the availability of co-workers would be shown instead.

Additionally, as \ap{} pointed out, not all data inherently has a spatial relationship with a physical referent, and thus it is not always straightforward to decide where to situate such data. In these cases, participants situated their dashboard design somewhere which they said would be most convenient for them to access. For example, \fleck{}, \beck{}, and \sr{} all indicated they would choose to place a dashboard near their work desk. For \fleck{} and \sr{} in particular, they emphasized the importance of the dashboard being within arm's reach for easy interaction, particularly if the dashboard contained many interactive controls. For similar reasons, \aimee{} suggested placing a dashboard on top of their students' desks to facilitate interactive teaching activities.

% In general, participants took the following aspects into account when situating the dashboard in context: relevance to physical referent, virtual or tangible proxy of physical referent, time of use, place of use, and convenience of interaction. 

\subsection{Interacting with the Dashboard}
When asked what modalities they would expect to use with situated dashboards, all participants suggested using mid-air hand, eye gaze, and tangible interactions. Voice input was not a preferred modality. \ap{} cited that it ``might be hard to use in noisy environments'', and \beck{} and \sr{} stated that it would likely be uncomfortable to use in public.

As previously mentioned, most participants would rather interact with dashboards that are within arm's reach, and would therefore avoid interacting with dashboards that were far away. In such a scenario, \sr{} said they would use the dashboard to only look at information, not interact with it. \beck{} instead said that gaze interaction on a distant dashboard could be a way to perform certain tasks. They proposed gazing at a calendar on the dashboard, which would then open it on their personal computer for them to make changes on it.

%Participants mostly did not prefer interacting with dashboards that are situated at a distance. However, \sr said that he might only use such dashboard to look at information or use as a shortcut or do simple tasks such as checking off boxes (\beck, \sr). For example, he mentioned looking at the calendar on the dashboard for a while as a way to open it up on his computer where he can make changes on the calendar.

Other participants described alternative methods for interacting with the dashboard. \fleck{} mentioned their dislike of mid-air interaction, suggesting that a mobile application or tangible slider could be used to manipulate the data and/or referent instead. \aimee{} similarly suggested that the dashboard could be aligned against a tabletop surface, with physical objects being used to interact with the dashboard.

When asked how to perform basic visualization tasks, \kadek{} proposed that data could automatically be filtered based on the physical proximity of the user to the referent. In contrast, \beck{} suggested that data could be manually filtered using checkboxes. They also envisioned using a pinch gesture to zoom into specific visualizations for more details, or by grab and dropping a visualization onto a secondary panel to expand it.
Following the same overview first, zoom and filter mantra \cite{shneidermanEyesHaveIt1996}, \fleck{} described a ``reactive situated dashboard'' which changes its level of detail via proxemics \cite{hallProxemicsCommentsReplies1968}. Alternatively, \fleck{} suggested that the user could \textit{``focus [their gaze] on something for an extended period of time, [...], it gets the information and you get more details on your dashboard.''}

% \beck, \fleck, and \aimee---all of them placed dashboards with interactive components within arm's reach and talked about interacting with them through hand interactions. However, \aimee and \fleck{} mentioned that they usually do not like mid-air hand interaction and so \fleck{} talked about using mobile app or physical slider to control temperature through dashboard and \aimee~ talked about positioning the dashboard on tabletop and using physical objects such as flashcards to interact with it.
% \subsubsection{Data filtering}
% The techniques for data filtering used by participants include automatic filtering of information based on proximity of user to the referent (\kadek), and manual filtering of information by tapping on checkboxes (\beck).
% \subsubsection{Overview-Zoom}
% \beck{} used pinching gesture to zoom in on specific visualizations. Other alternative approaches he suggested for overview-zoom were using a second panel where he can grab and drop a visualization from the dashboard to reveal details. 

% \fleck{} suggested using "reactive situated dashboard" which does overview-detail when user comes closer to it. Additionally, he mentioned a strategy where if users "focus on something for an extended period of time, which is unusual to [...] routine while wearing AR glasses, [...] it gets the information and you get more details on your dashboard".

\subsection{Customizing the Dashboard}
\kadek{} and \beck{} emphasized the need to provide customization support for end-users to personalize their experiences with their situated dashboard. \beck{} suggested providing ``building blocks'' so that end-users \textit{``can build a solution that they need.''}
However, \beck{} also acknowledged the limitations of using building blocks to author entire situated dashboards. While it may be relatively easy to provide simple means to customize the layout of the dashboard, for example, they noted that ``\textit{[considering the] whole situated thing and like the context switching and so on [...] it becomes a lot more complicated.}''
\section{Challenges}
We now discuss several challenges associated with situated dashboards. Most of them are based on the interviews, and others are based on our own internal discussions. Note that some challenges apply to the broader subject of situated visualization and analytics.

\subsection{C1: (Situated) Authoring of Situated Dashboards}
Despite it not being a common discussion topic in our interviews, we believe that the authoring process of situated dashboards is an obvious next step and research challenge.

An important consideration is the level of expertise expected of the end-user. At present, a small number of situated analytics toolkits exist---most notably, RagRug \cite{fleckRagRugToolkitSituated2022}. When talking about their typical workflow for implementing situated visualization with RagRug, \fleck{} stated that it was easy to use while \aimee{} firmly stated it was not. This disagreement between our own participants suggests the need for situated analytic toolkits that are easier for novices to use.

We believe this need is exacerbated when considering the potential end-users of such a toolkit. While most related work has considered situated analytics in some specific domain (e.g., building maintenance~\cite{prouzeauCorsicanTwinAuthoring2020}, sports~\cite{linUnderstandingSituatedAR2021}), we speculate that situated dashboards could be used in any context that involves data. Rather than devising a ``one size fits all'' application, end-users with limited expertise may want or need to customize and/or personalize their situated dashboards to suit their goals, data sources, and physical environments. Consider a restaurant manager who wants to have an AR situated dashboard to keep track of stock levels. Instead of hiring an expert to create the dashboard, the manager wants to do it by themself to properly tailor it to their own preferences and needs. The toolkit therefore needs to be simple enough for even laypersons to use, but be expressive enough to have utility in a wide range of scenarios.

The best authoring paradigm however is unclear. In broader immersive analytics, authoring systems range from text-based specifications (e.g.,~\cite{DXR2019,butcherVRIAWebBasedFramework2021}) to GUIs (e.g.,~\cite{cordeilIATKImmersiveAnalytics2019}) to fully embodied interactions (e.g.,~\cite{cordeilImAxesImmersiveAxes2017}). The latter approach would likely involve ``building blocks'', as \beck{} suggested, to allow end-users to easily build situated dashboards without complex grammars or code. Other researchers have also suggested this approach \cite{leeDeimosGrammarDynamic2023}, but it can limit expressiveness if not enough presets and templates are provided.
That said, if situated dashboards were instead used as interaction panels as per \aimee{} and \sr{}, then complex visualization toolkits may not even be needed.

\sr{} and \aimee{} had expressed frustrations in creating AR visualizations. At present, deployment requires a switch between development and situated contexts, incurring a high temporal and cognitive cost. Tools like Corsican Twin~\cite{prouzeauCorsicanTwinAuthoring2020} circumvent this by allowing authoring of situated visualizations immediately in the physical environment itself. This form of situated authoring would likely be ideal for creating situated dashboards in the future.
Situated authoring may also serve to explicitly connect and link the data of referents to the dashboard's visualizations. Ivy by Ens et al.~\cite{ensIvyExploringSpatially2017} demonstrates this idea by using 3D visual links to connect data nodes in a 3D environment together. While certainly straightforward, such direct linking might not be practical when referents are either too far away, too high in number, or are not spatially registered. RagRug~\cite{fleckRagRugToolkitSituated2022} provides a more standardized solution to link data sources to visualizations via MQTT, but this approach may be too technical for laypeople. Thus, finding an appropriate solution for this would be paramount for situated dashboards (and visualization as a whole).

The concept of context-awareness came up numerous times in our interviews. The dashboard may change and adapt depending on contextual factors, such as changing views based on the user's spatial proximity to referents. The challenge here is not only ensuring the system itself is context-aware \cite{baldaufSurveyContextawareSystems2007}, but also to investigate how the end-user might best define dashboard adaptations based on their chosen contextual factors.

\subsection{C2: Dashboard Layout \& Scalability}
The choice of dashboard layout may be challenging as this influences its effectiveness.
The standard approach would be to use 2D dashboards as floating panels. Their similarity to conventional dashboards may prove to be their strength, and all participants only considered 2D visualizations in our interviews.
In contrast, no participants mentioned using 3D visualizations at all, which is unsurprising given their perception issues and propensity to occlude. \kadek{}, however raised an interesting point in that a 3D proxy comprised of multiple referents may function as a dashboard (i.e., proxsituated visualization \cite{proxSituated2023}). While the proxy would look like a 3D world-in-miniature, its purpose would serve mostly as a 3D overview of the full environment rather than for navigation or manipulation \cite{danylukDesignSpaceExploration2021}. It may be that while a 2D dashboard provides a familiar overview of the data, a 3D dashboard may perform better when understanding the spatial layout of the data and referent is paramount.

The question of scalability also arose in our interviews. \fleck{} suggested that having too many visualizations on a dashboard would necessitate some form of filtering. This filtering may be automatic based on context or be performed manually by the end-user. Alternate representations of data may also need to be employed. Rather than one visualization per referent, all referents could be aggregated into a single one. The trade-off however is that it may unintentionally hide important information. A third approach may be to embrace the large number of visualizations. As immersive devices are oftentimes touted by their ability for large workspaces, dashboards could be infinitely scaled to present large amounts of data. While this may obscure the surrounding environment in an AR context, a cross-virtuality setup could be employed to transition the end-user into VR, resulting in a ``focused'' mode to analyse the dashboard's data. How best to handle this scalability issue remains unclear. 

% - 2D vs 3D dashboards (dimensionality)
%     - Only comprised of 2D visualisations
%         - Similar to a regular dashboard
%         - Familiar and recognisable setup
%     - Only comprised of 3D visualisations
%         - A compelling example is a form of a WiM proxy of the physical referent
%         - Imagine a smaller scale model of a factory with labels on different parts of the factory
%     - Comprised of both 2D and 3D
%         - Layout might be a concern
%         - Position vertically vs flat horizontally, but 3D vis might occlude other vis
%         - How to coordinate multiple visualizations within the dashboard?.
% - How does a dashboard accommodate lots of referents (scalability)?
%     - Filtering would likely be needed
%         - Filtering based on spatial position, context, time, manual filtering
%     - Even higher level aggregation of data
%         - But this might hide away some important information
%     - Embrace the chaos, take advantage of the large space around the user
%         - Can obscure the surrounding environment
%         - But fully immerse the user in the data
%         - Could be beneficial in a cross-reality setup, i.e. transition to VR for a "focused" mode of analysing the dashboard

\subsection{C3: Placement and Interaction of Dashboards}
From our interviews, the placement of situated dashboards depends on the data and the end-user's intention to interact with it. While it might be imperative to place the dashboards in places where it provides actionable information to users (i.e., nearby the referent), many participants preferred the dashboard to be at arm's reach to make interaction easier. Even so, arm's reach may require the dashboard to float in mid-air, or be overlaid against a wall or table to enable touch-like input. This demonstrates a challenge in balancing between proper situatedness of the dashboard, and ease of interaction regardless of the end-users physical proximity to the referent.

Interestingly however, no participants talked about how a situated dashboard might move with the end-user throughout the physical environment, even though proximity to referents was the main example given for context-aware dashboards. It is safe to assume that dashboards could be moved manually, but automatic solutions may also be employed (e.g., \cite{evangelistabeloAUITAdaptiveUser2022}). However, dashboards can vary greatly in terms of their size, content, and appearance. It might even be imperative that a specific dashboard be placed next to its referents, acting as a hard requirement for its placement. Future work may consider these factors and determine how best to address them.

% - This is basically view management
% - How to place the dashboard?
%     - Offer automatic solutions (e.g. AUIT)
%     - Offer fixed solutions (body fixed, world fixed, surround fixed)
%     - Attached to physical objects (but risk occlusion)
%     - Responsiveness
% - How to place it to facilitate interaction?
%     - Requiring the user to walk vs having it in arms reach
%     - Moving the dashboard vs moving the person
%     - Minimize/maximize
% - How to share, save, export dashboards (collaboration)

\subsection{C4: Navigation between Dashboard and Referent}
\ap{} briefly described how visualizations on a situated dashboard could be associated with referents. This can be considered as an overview first, zoom and filter interaction \cite{shneidermanEyesHaveIt1996}. The end-user identifies a referent on the dashboard, then navigates to its physical location which may contain other situated or embedded visualization(s). This may require a form of visual guidance by the system. If the referent is in close proximity, simple attention guidance like a visual link is enough (e.g., \cite{prouzeauVisualLinkRouting2019}). If further away, a more complete navigation technique might need to be employed instead (e.g., \cite{reitmayrCollaborativeAugmentedReality2004,mulloniHandheldAugmentedReality2011})).
Willett et al.~\cite{willettEmbeddedDataRepresentations2017} suggested that visualizations could transition from non-situated to situated to embedded. Thus, an interesting consideration is whether a transition occurs between the dashboard and any situated/embedded visualizations on the referent. If both dashboard and referent visualizations use the same idiom, then this is trivial: use the same visualization. But if they use different idioms, then designing a suitable transition may prove challenging.

% - How do we link between visualisation and referent?
%     - If in close proximity, something like a visual link is enough
%     - If further away, a more complete navigation technique might be required
%     - To avoid navigation, standard data from the referent, must be represented by common visualizations. (Virtual Proxy?)
% - Are links even necessary?
%     - A proxy of the referent is technically enough to understand its physical context
%     - However, it obviously means that you cannot interact with the actual physical context
% - How does the transition move between dashboard and visualisation when in spatial proximity?
%     - Implicit overview + zoom can filter the dashboard to the data on the nearby referent
%     - But what if the design pattern used for both are incompatible (e.g. scatterplot on dashboard, but decals on the referent?). This is a challenge

\subsection{C5: Moving in/out/between Situated Environments}
While our interviews discussed what makes a situated dashboard ``situated'', an interesting consideration is what happens when the user moves in, out, or between configured situated environments. Consider someone walking into a store. Does a dashboard of store prices suddenly appear in front of them, or is it fixed near the store's entrance? Now consider the same person moving to another store, with it also having its own situated dashboard application. Does the same dashboard change content, does a new dashboard appear and replace the other, or do multiple dashboards appear simultaneously?
These questions relate to the broader societal context surrounding each situated environment. If situated dashboards become ubiquitous and are loaded on-the-fly as we move about the physical world, how does the system manage each situated environment? Who decides who ``owns'' a particular spatial region in which a dashboard or visualizations appears in? For situated analytics to become commonplace, these questions likely need to be addressed.

% - The dashboard moving away from its physical context:
%     - Is it no longer considered as situated, and is this a problem?
%     - Does it disappear or does it remain?
%     - Should situated-dependent functions like overview + zoom or remote interaction still function when the dashboard is no longer considered situated?
% - The user physically moving from one environment that has a configured situated dashboard to another:
%     - Do they see multiple dashboards at a time?
%     - When multiple situated dashboard creators overlap in their spatial regions, who gets ownership or priority?

\subsection{C6: Collaborative and Remote Situated Evironments}

Given that AR allows users to interact with real scenarios and permits direct communication, collaborative and remote approaches must be studied~\cite{grandChallenges2021}. Our proposal about situated dashboards is not limited to collaborative and remote tasks of conventional authoring tools. We go beyond proxsituated visualizations~\cite{proxSituated2023} and envision remote authoring situated dashboards. Although tangible interaction performs helpful in some scenarios, the collaborative work would get huge benefits by sharing non-accessible non-reproducible referents.

\section{Conclusion and Future Work}
In this position paper, we proposed and investigated the concept of situated dashboards. We identified key design considerations and challenges through interviews with six researchers in (situated) visualization and/or AR. It is apparent that there is no singular agreed upon definition of situated dashboards, let alone their design, behavior, and interactivity. Thus, future work on situated dashboards and analytics as a whole may need to support a wide range of different designs and contexts, in order to support the needs of a wide range of end-users. The challenges we identified thus serve as future research directions in service of this broader goal.

\acknowledgments{
We would like to thank Samuel Beck, Aimee Sousa Calepso, Philipp Fleck, Arnaud Prouzeau, Sebastian Rigling, and Kadek Satriadi for participating in the study. This research was funded by the German Research Foundation (DFG) project 495135767 and the Austria Science Fund (FWF) project I 5912-N (joint Weave project). The project is associated with and further supported by the DFG Excellence Cluster EXC 2120/1 – 390831618.}

\bibliographystyle{abbrv-doi}

\bibliography{template}
\end{document}